 \documentclass[final,5p,times,twocolumn,authoryear]{elsarticle}
\usepackage{amssymb}
\usepackage{amsmath}
\usepackage{color}
\def\psrb{PSR B1259-63/LS 2883}

\def\gammaray{$\gamma$-ray}
\def\gammarays{$\gamma$-rays}
\def\gaga{$\gamma\gamma$}
\def\hess{H.E.S.S.}

\def\fermi{\emph{Fermi}-LAT}
\def\deg{$^{\circ}$}

\newcommand{\IS}[1]{#1} %\textcolor{red}{\textbf{\boldmath#1\unboldmath}}}
\newcommand{\IStwo}[1]{#1} %{\textcolor{red}{\textbf{\boldmath#1\unboldmath}}}
\newcommand{\newref}[1]{#1}%{\textcolor{red}{\textbf{\boldmath#1\unboldmath}}}

\journal{Journal of High Energy Astrophysics}

\begin{document}

\begin{frontmatter}



\title{Pair Cascades in the Disk Environment of the Binary System \psrb.}


\author[nwu,lviv]{Iurii Sushch}
\author[nwu,ohio]{Markus B\"ottcher}
\address[nwu]{Centre for Space Research, North-West University, Potchefstroom 2520, South Africa}
\address[lviv]{Astronomical Observatory of Ivan Franko National University of L'viv, vul. Kyryla i Methodia, 8, L'viv 79005, Ukraine}
\address[ohio]{Astrophysical Institute, Department of Physics and Astronomy, Ohio University, Athens, OH 45701, USA}
\begin{abstract}
\psrb\ is a very high energy (VHE; $E>100$ GeV) $\gamma$-ray emitting binary consisting of a 48 ms pulsar orbiting around a Be star with a period of $\sim3.4$ years. The Be star features a circumstellar disk which is inclined with respect to the orbit in such a way that the pulsar crosses it twice every orbit. The circumstellar disk provides an additional field of target photons which may contribute to inverse Compton scattering and \gaga-absorption, leaving a characteristic imprint in the observed spectrum of the high energy emission. At GeV energies, the source was detected for the first time during the previous periastron passage which took place on December 15, 2010. The \emph{Fermi} Large Area Telescope (LAT) reported a spectacular and unexpected \gammaray\ flare occurring around 30 days after periastron and lasting for about 7 weeks. In this paper, we study the signatures of Compton-supported, VHE \gammaray\ induced pair cascades in the circumstellar disc of the Be star and their possible contribution to the GeV flux. 
\newref{We show that cascade emission generated in the disk cannot be responsible for the GeV flare, but it might explain the GeV emission observed 
close to periastron. We also show that the \gaga-absorption in the disk might explain the observed TeV light curve.}
\end{abstract}


\end{frontmatter}


\section{Introduction}
\label{intro}
\psrb\ is a member of the small class of very high energy (VHE; $E>100$ GeV) \gammaray\ binaries which comprises only 
five known objects. \psrb\ is unique for being the only \gammaray\ binary for which the 
compact object is clearly identified as a pulsar. This pulsar with a spin period of $\simeq 48$~ms and a spin-down 
luminosity of $\simeq 8 \times 10^{35}$~erg~s$^{-1}$ is moving in a highly eccentric ($e = 0.87$) 
orbit around a massive Be star with a period of $P_{\mathrm{orb}} = 3.4$ years (1237 days) 
\citep{1992MNRAS.255..401J, 1992ApJ...387L..37J}. The companion star LS 2883 has a luminosity 
of $L_{\ast} = 2.3 \times 10^{38}$~erg~s$^{-1}$. Because of the fast rotation of the star, it has an oblate shape 
with an equatorial radius of $R_{\mathrm{eq}} = 9.7 R_{\odot}$ and a polar radius of $R_{\mathrm{pole}} = 8.1 R_{\odot}$, 
which in turn leads to a strong gradient of the surface temperature from $T_{\mathrm{eq}} \simeq 27,500$\,K at the equator 
to $T_{\mathrm{pole}} \simeq 34,000$\,K at the poles \citep{2011ApJ...732L..11N}. 

The Be star features an equatorial 
circumstellar disk, which is believed to be inclined with respect to the pulsar's orbit 
\citep[see e.g.][]{2011ApJ...732L..11N}, so that the pulsar crosses the 
disk twice each orbit. The circumstellar disk of a Be star is a decretion disk with an enhanced stellar outflow formed around the star. 
As shown in optical interferometry observations, these disks are symmetrical with respect to the star's rotation 
axis \citep[see e.g.,][]{1994A&A...283L..13Q}. 
Circumstellar disks generate excess infrared (IR) emission produced through free-free and 
free-bound radiation, providing an additional IR photon field to the blackbody flux from the optical star. 
%which may significantly contribute to the inverse Compton scattering.

The dense medium of the disk is believed to play an essential role in the resulting emission from the system. The 
position of the disk can be localized based on the disappearance of the pulsed radio emission from the pulsar. 
The observed radio emission far from periastron consists only of the pulsed component \citep{1999MNRAS.302..277J, 2005MNRAS.358.1069J}, 
but closer to periastron, at about $t_{\mathrm{p}} - 100$\,d ($t_{\mathrm{p}}$ is the time of periastron), the intensity of the 
pulsed emission starts to decrease and completely disappears at about $t_{\mathrm{p}} - 20$\,d. It then re-appears at around $t_{\mathrm{p}} + 15$\,d. 
This eclipse of the pulsed emission is believed to be caused by the circumstellar disk.
% the pulsed signal disappears as the pulsar goes behind the 
% disk. The eclipse
It is accompanied by an increase of the transient unpulsed radio flux beginning at 
$\sim t_{\mathrm{p}}-30$\,d and reaching its maximum at $\sim t_{\mathrm{p}}-10$\,d. This is followed by a decrease around the 
periastron passage and a second peak at about $t_{\mathrm{p}} + 20$\,d \citep{1999MNRAS.302..277J, 2005MNRAS.358.1069J, 2014MNRAS.439..432C}.
A similar behavior is observed also for the unpulsed X-ray emission. Close to the periastron passage the X-ray emission features 
two peaks at around 20 days before and after periastron with flux levels 10 -- 20 times 
higher than during apastron, and a decrease of the emission at the time of the periastron passage itself. 
The X-ray data is very similar from orbit to orbit repeating the shape of the light curve 
very well \citep[][and references therein]{2014MNRAS.439..432C}. These peaks might be connected 
to the crossing of the disk environment.

At TeV energies the source was observed by \hess\ around three periastron passages in 2004 \citep{2005A&A...442....1A}, 2007 
\citep{2009A&A...507..389A} and 2010 \citep{2013A&A...551A..94H}. The TeV emission from the source shows a variable behavior 
around the periastron passage. Although the exposure of the source is not sufficient to draw firm conclusions, the combined 
light curves from all three observing campaigns show a hint of two asymmetrical peaks before and after periastron, which 
roughly coincide in time with the peaks of the emission in the radio and X-ray bands, and a decrease of the flux at periastron. 
This enhancement of the flux before and after periastron might be induced by the interaction of the pulsar with the dense 
photon field region inside the circumstellar disk.

First observations at GeV energies conducted by \fermi\ took place around the 2010 periastron passage revealing 
quite unexpected results \citep{2011ApJ...736L..11A, 2011ApJ...736L..10T}. \fermi\ detected a low flux from the source 
close to periastron with a subsequent dissapearance of the 
source after periastron followed by a sudden flare (with $\sim10$ times the pre-periastron flux level) 30 days after periastron. 
This flare lasted for about 7 weeks without any obvious counterparts at other energy bands. The flare is shifted in time 
with respect to the post-periastron peak at other energy bands. The nature of the flare is still not understood, but 
several possible explanations have been discussed in the literature. \citet{2012ApJ...752L..17K} proposed that the GeV flare might 
be the result of inverse Compton (IC) scattering of stellar and circumstellar disk photons by the unshocked pulsar wind. 
The model suggests that right after the pulsar escapes the circumstellar disk, the sharp decrease of the ram pressure rapidly 
increases the unshocked pulsar wind zone towards the observer which, assuming that the density of the seed photons is still 
high, results in the observed GeV flare. The enhancement of the GeV emission is not observed in the pre-periastron phase due to an 
unfavourable geometry. Another possible explanation of the GeV flare can be the Doppler boosting of the radiation created 
by the shocked pulsar wind \citep{2008MNRAS.387...63B, 2010A&A...516A..18D, 2012ApJ...753..127K}. However, this scenario is 
problematic because the Doppler boosting should affect the emission in all energy bands, but no counterparts of the GeV flare at 
other energy bands were detected. \citet{2012ApJ...753..127K} tried to solve this problem by assuming a specific anisotropy of the 
pulsar wind with different emission behaviors in different regions of the termination shock. \citet{2013A&A...557A.127D} 
suggested that the GeV flare may be generated by IC scattering of the X-ray photons from the pulsar wind nebula instead of the 
stellar photons. In this model, the lightcurve naturally peaks after the periastron as the cone of shocked material passes 
through the line of sight. However, the model does not explain the delay between the GeV flare and the post-periastron X-ray peak. 
It should be noted that all the proposed models suggest that the flare occurs due to geometrical effects

The circumstellar disk of the companion star plays a crucial role in the variability of the \gammaray\ emission from 
the system. The dense disk photon field should significantly contribute to the target photons for inverse Compton scattering from the source, enhancing 
the observed TeV emission. The abrupt change of density at the pulsar's entrance and escape from the disk should also cause 
a change of the shape of the shock between the pulsar wind and stellar environment, which might be the reason of such spectacular 
events as the GeV flare. The high density of seed photons should also increase the opacity for \gaga-absorption, followed by electron-positron 
pair production, which in turn might scatter again the disk photon field, generating secondary \gammarays. This cascade 
process may cause the re-emission of the TeV IC flux at lower GeV energies. Moreover, because of the deflection of electrons 
and positrons in the magnetic field, secondary \gammarays\ can be re-emitted into randomized directions. Therefore, even if 
the primary VHE photon was emitted into the direction opposite to the line of sight to the observer, it can still contribute to the observable 
flux at lower GeV energies through the cascade emission. The pair cascade emission in binary systems caused by the interaction 
of the primary very high energy photons with the stellar photons was studied in detail for several cases 
\citep[see e.g.][and references therein]{2005MNRAS.356..711S, 2010A&A...519A..81C}. These studies showed that, although environments of the binary systems 
fulfill all the requirements for effective pair cascading, the resulting spectrum is in conflict with the GeV data observed by \fermi, 
since the expected cascade emission peaks in the \fermi\ energy band with very constraining upper limits. The \gaga-absorption of the 
TeV \gammarays\ by stellar photons was also suggested as a possible explanation of the variability of the TeV flux across the orbit 
\citep{2006A&A...451....9D}. The idea is that at periastron, when the pulsar is at the shortest distance to the star, the absorption 
should be the most effective, which would provide the decrease of the flux. However, \citet{2006A&A...451....9D} showed that absorption only 
cannot explain the TeV light curve.

In this study we investigate 
the possibility of pair cascading in the binary system \psrb\ induced by the interaction with the circumstellar disk photons alone without 
taking into account the stellar wind from the companion star. We numerically evaluate the contribution of the cascade emission produced in 
the circumstellar disk to the overall spectrum. The comparison of the resulting modified spectrum with the observational data may be used to probe 
the circumstellar disk environment.

\section{Model Setup}
\label{model}
\begin{figure}
\centering
\resizebox{\hsize}{!}{\includegraphics{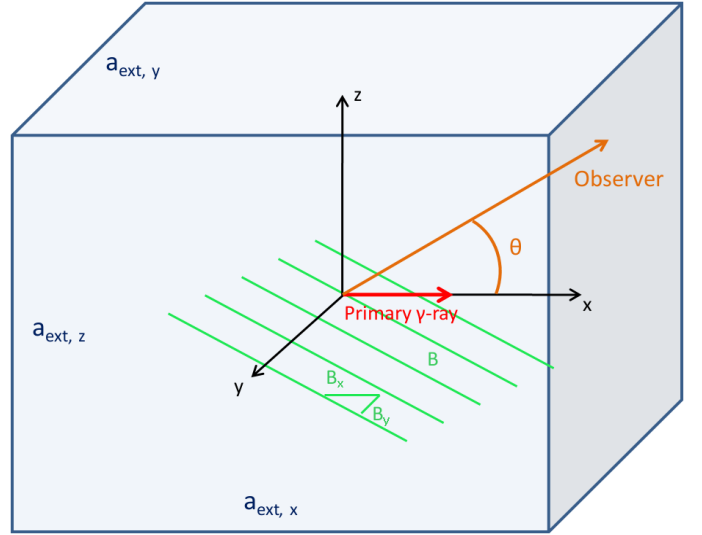}}
\caption{Geometry of the model setup}
\label{geom_model}
\end{figure}

The geometry of the model setup is illustrated in Fig.\,\ref{geom_model}. In order to be able to isolate all geometric effects, we represent the primary VHE emission as a 
mono-directional beam and define the \emph{x}-axis as the direction of propagation of the primary VHE $\gamma$-rays. 
The VHE \gammaray\ spectrum is described 
by a power law with an exponential cut-off,
\begin{equation}
  N(E_{\gamma}) \propto E_{\gamma}^{-\alpha} \exp(-E_{\gamma}/E_{\mathrm{cut}}),
\end{equation}  
where $E_{\gamma}$ is the photon energy, $\alpha$ is the spectral index, and $E_{\mathrm{cut}}$ is the cut-off energy. For 
an initial, general study we assume that the primary \gammarays\ interact via \gaga-absorption and pair production with the disk 
photons, approximated by an isotropic thermal blackbody radiation field within fixed spatial boundaries given by a cuboid with 
side lengths $a_{\mathrm{d,\,x}}$, $a_{\mathrm{d,\,y}}$ and $a_{\mathrm{d,\,z}}$, i.e.,
\begin{align}
 u_\mathrm{d}(\nu, \mathbf{r}, \mathbf{\Omega}) =
  \begin{cases}
    \frac{2h\nu^3}{c^3} \frac{A}{\exp(\frac{h\nu}{kT}) - 1}& \text{if } |x| \leq \frac{a_{\mathrm{d,\,x}}}{2}, |y| \leq \frac{a_{\mathrm{d,\,y}}}{2}, |z| \leq \frac{a_{\mathrm{d,\,z}}}{2} \\
   0       & \text{if } |x| > \frac{a_{\mathrm{d,\,x}}}{2}, |y| > \frac{a_{\mathrm{d,\,y}}}{2}, |z| > \frac{a_{\mathrm{d,\,z}}}{2}
  \end{cases}
\end{align}
where $\nu$ is the target photon frequency, $h$ is the Planck constant, $k$ is the Boltzman constant, $c$ is the speed of light, $T$ is the temperature and $A$ is a normalization factor chosen to obtain a total radiation energy density $u_{\mathrm{d}}$ 
%% (see Eq. \ref{u_ext} below) 
\begin{equation}
\label{u_ext}
  u_{\mathrm{d}} = 4\pi \int_0^{\infty} u_{\mathrm{d}}(\nu, \mathbf{r}, \mathbf{\Omega}) d\nu.
\end{equation}
We choose the \emph{y-} and \emph{z-}axes of the coordinate system in 
a way that the magnetic field \textbf{B} lies in the \emph{xy-}plane.

For our cascade simulations, we have adapted the Monte Carlo code developed by \citet{2010ApJ...717..468R, 2012ApJ...750...26R} 
to our chosen disk geometry. The code calculates \gaga-absorption and 
pair production using the full analytical solution to the pair production spectrum of \citet{1997A&A...325..866B} under 
the assumption that the produced electron and positron travel along the direction of propagation of the incoming VHE photon. 
The trajectories of particles are followed in full three-dimensional geometry. Inverse Compton scattering is 
evaluated using a head-on approximation, assuming that the scattered photon travels along the direction of motion of the 
electron/positron at the time of scattering. Compton and synchrotron losses are properly taken into account. Tables for 
the absorption opacity $\kappa_{\gamma\gamma}$, Compton scattering length $\lambda_{\mathrm{IC}}$ and Compton cross section 
for dense grids of photon energy, electron energy, and interaction angle are pre-calculated before the start of simulations, in order 
to save CPU time. The simulation produces a photon event file containing the energy, statistical weight, and propagation 
direction of every photon which escapes the cuboid target photon region. A separate post-processing routine is used to 
extract the angle-dependent photon spectra with arbitrary energy and angular binning. 

\section{Application to \psrb}
\label{psrbapp}

\IS{\subsection{Model Assumptions}
\label{assump}
\psrb\ is a rather complicated source for modeling due to the variety of different physical processes which 
may take place in the system. The binary system consists of the pulsar and the massive Be star which produces a 
stellar wind and features a dense equatorial circumstellar disk. The interaction of the pulsar wind with the 
stellar wind and/or circumstellar disk of the companion star results in the creation of a termination shock 
between the two winds. Relativistic leptons (electrons and positrons) injected by the pulsar can be accelerated and 
isotropised at the termination shock and subsequently interact with the stellar radiation fields (stellar wind and 
circumstellar disk) producing very-high-energy \gammarays\ via inverse Compton scattering. These primary 
\gammarays\ can be absorbed in the stellar radiation fields producing electron-positron pairs, which in turn 
propagate in the stellar magnetic field and further scatter stellar and circumstellar 
disk photons producing the second generation of \gammarays. This cascade will continue until the lepton does not 
have enough energy to produce another secondary photon and/or the \gammaray\ escapes from the system. To provide a 
first general study of the pair cascades generated in the disk environment we will use several assumptions to 
constrain the stellar-pulsar wind interaction, the stellar radiation field, and the magnetic field structure. 
}

%% Point-source assumption justification
\IS{The location of the wind termination shock can be estimated from the winds' ram pressure ratio 
$\eta  = L/(\dot{M_\ast}c\upsilon_\ast)$ \citep{2000APh....12..335B}, where $L$ is the pulsar spin-down luminosity and $\dot{M_\ast}$ 
and $\upsilon_\ast$ are the stellar wind mass-loss rate and the stellar wind velocity at the interaction 
position. Wind characteristics are different in the polar and equatorial regions. The polar stellar wind 
has a relatively low density and high velocity, while the equatorial wind is much denser and slower. 
This difference results in the different geometry of the wind interaction in different orbital phases. 
For known $\eta$ the distance from the pulsar to the termination shock is given by 
\begin{equation}
  \rho = d \frac{\sqrt{\eta}}{(1+\sqrt{\eta})},
\end{equation}
where $d$ is the separation distance between the star and the pulsar. According to \citet{2000A&A...362..295V} 
the mass and effective temperature of LS 2883 correspond to the 
mass-loss rate of $(0.6 - 1.5) \times 10^{-7} M_{\odot}/\mathrm{yr}$ (in the 
polar region) depending on the assumed vaule of the ratio 
$\upsilon_\infty / \upsilon_{\mathrm{esc}} = 2.6 - 1.3$ respectively, 
where $\upsilon_\infty$ is the terminal velocity of the wind at infinity 
and $\upsilon_{\mathrm{esc}}$ is the effective escape velocity. 
In the polar region the wind velocity can be aproximated by the 
profile 
\begin{equation}
\upsilon_{\mathrm{p}}(r) = \upsilon_0 + (\upsilon_\infty - \upsilon_0)(1 - \frac{R_\ast}{r})^{\beta},
\end{equation}
where $r$ is the distance from the star. 
Assuming $\upsilon_0 \simeq 10$\,km\,s$^{-1}$ and $\beta = 1.5$ \citep[see e.g.][and references therein]{2008MNRAS.385.2279S}, 
$\upsilon_\infty = 1350\pm200$\,km\,s$^{-1}$ \citep{1993AAS...183.1805M}, and the separation distance at 
periastron of $23R_\ast$, the polar wind velocity is $\upsilon_{\mathrm{p}} = 1260$\,km\,s$^{-1}$. Then the 
$\eta$ parameter for the periastron passage is  $\eta = (3 - 8) \times 10^{-2}$ and the distance to the 
termination shock is equal to $\rho = (0.15 - 0.22)d = (3.4 - 5.1)R_\ast$. When the pulsar crosses the 
circumstellar disk the ram pressure ratio will be much smaller due to the much higher value of the mass-loss 
rate, which in equatorial region is about $\dot{M_\mathrm{d}} = (10^2 - 10^4)\dot{M_\mathrm{p}}$ 
\citep[][and references therein]{2008MNRAS.385.2279S}. The equatorial wind velocity described by 
\begin{equation}
\upsilon_{\mathrm{d}}(r) = \upsilon_0(\frac{r}{R_\ast})^{1.25} \simeq 1200\mathrm{km\,s}^{-1}
\end{equation}
at the corresponding distance of $45R_\ast$ (the separation distance at $\sim 20$ days after periastron 
\citep[see e.g.][]{2012MNRAS.426.3135V} is, however, comparable to the one at periastron. Therefore, in 
the disk $\eta$ is $(10^2 - 10^4)$ lower than at periastron resulting in the distance to the 
termination shock of $\rho \sim (0.001 - 0.01)d = (0.05 - 0.5)R_{\ast}$. Moreover, the inclination angle 
of the equatorial disk to the orbital plane is argued to be rather small, at about $\sim10$\deg\ \citep[][see below]{1995MNRAS.275..381M}, i.e. 
even at periastron the environment should be influenced by the dense equatorial wind resulting in a lower 
value of $\eta$. Therefore, we decide that for our simulations the source can be assumed point-like neglecting 
the distance from the pulsar to the termination shock, where the electrons and positrons are accelerated.
}

%% Justification of not taking stellar photons into account
\IS{Both stellar raditation fields (stellar wind and circumstellar disk) are expected to contribute to the 
Comptonization by relativistic leptons. However, the contribution of these two radiation fields is different 
in different regions of the system and different orbital phases, i.e. the location of the pulsar within the 
system. The energy density of the stellar radiation can be estimated as 
\begin{equation}
 u_\ast(r) = \frac{L_\ast}{4\pi r^2 c}.  
\end{equation}
This corresponds to $u_{\ast,\mathrm{p}} \simeq 2.8$\,erg\,cm$^{-3}$ in the vicinity of the pulsar when the pulsar is 
at periastron and to $u_{\ast,\mathrm{d}} \simeq 0.7$\,erg\,cm$^{-3}$ when the pulsar is in the disk (20 days after 
periastron). If the energy density of the disk photons is higher than the stellar photon energy density in the same 
region, we can neglect the contribution of the stellar photons to the \gaga-absorption and to the IC scattering, and 
consider circumstellar disk radiation as the only source of target photons. The binary separation at $\sim 20$ days provides 
a lower limit for the disk radius $R_{\mathrm{d}}$ of $\sim 45 R_{\ast}$, where radius of the star is $R_{\ast} = 9.2 R_{\odot}$ 
\citep{2012MNRAS.426.3135V}, i.e. $R_{\mathrm{d}} \gtrsim 2.9\times10^{13}$ cm. Circumstellar disks of Be stars are believed 
to be very thin \citep[see][and references therein]{2013A&ARv..21...69R}. In recent theoretical works on \psrb, a value of 
$0.7$\deg$-1$\deg\ was adopted for the disk half-opening angle $\theta_{\mathrm{d}}$ \citep{2011PASJ...63..893O, 2012ApJ...750...70T, 2012MNRAS.426.3135V}, 
which corresponds to a width of $1 \times 10^{12}$ cm at the distance of $45 R_{\ast}$. 
The fit of the dispersion and rotation measures suggests a small value of the inclination angle of the disk with 
respect to the orbit of about $10$\deg\ \citep{1995MNRAS.275..381M}. Although this fit requires a quite high 
value of the magnetic field strength at the star surface of $B_{\mathrm{S}} = 14$ kG \citep{2012SSRv..166..145W}, 
larger disk inclincation angles require an even stronger magnetic field. Therefore the inclination angle is expected 
to be small $\lesssim10$\deg. However, it should be noted that other values, such as $45$\deg\ and $90$\deg, 
were also considered in theoretical modeling of the system by different authors \citep[see e.g.][]{2011PASJ...63..893O, 2012MNRAS.426.3135V}. 
Assuming the inclination angle of the disk is $10$\deg\ and the disk half-opening angle is $1$\deg, the path length 
of the pulsar inside the disk at the distance of $45R_\ast$ from the star would be $\sim6\times10^{12}$ cm. 
A temperature of the disk of $T_{\mathrm{d}} = 0.6 \, T_{\ast}$ is usually assumed \citep{2011PASJ...63..893O, 2012ApJ...750...70T, 2012MNRAS.426.3135V}, 
which corresponds to $T_{\mathrm{d}} = 18,000$ K for $T_{\ast} = 30,000$ K. Here we approximate the disk by a grey-body with the energy density 
$u_{\mathrm{d}}$ and temperature $T_{\mathrm{d}}$, the shape of the disk region we approximate by a cuboid 
with side lengths $a_{\mathrm{d,\,x}} = 5\times10^{12}$ cm, $a_{\mathrm{d,\,y}} = 1\times10^{13}$ cm, 
and $a_{\mathrm{d,\,z}} = 1 \times 10^{12}$ cm. The value of $u_{\mathrm{d}}$ is constrained by the total lumonosity of the 
star $L_\ast$, i.e.
\begin{equation}
u_{\mathrm{d}}<\frac{4L_\ast}{c S_{\mathrm{d}}}\simeq 200\,\mathrm{erg\,cm}^{-3}, %% 236
\end{equation}
where $S_{\mathrm{d}}$ is the effective surface of the disk. In the following, we  assume 
the circumstellar disk energy density in the range of 
\begin{equation}
0.7\,\mathrm{erg\,cm}^{-3} < u_{\mathrm{d}}< 200\,\mathrm{erg\,cm}^{-3},
\end{equation}
which allows us to neglect the contribution of the stellar photons to the considered physical 
processes.
}

%% Justification of the magnetic field
\IS{
The magnetic field in the wind of the massive star has a complicated structure which 
depends on the distance from the star \citep[see e.g.][and references therein]{2005MNRAS.356..711S}. 
In the region very close to the stellar surface the magnetic field has a dipolar structure. At 
a distance characterized by the Alfven radius $R_{\mathrm{A}}$ the radial component of the 
magnetic field starts to dominate due to the ionized plasma. Finally, at larger distances, the magnetic field 
becomes toroidal due to the rotation of the star. The strength of the magnetic field as a function of 
distance from the star is given by \citep{1992ApJ...395..575U}
\[B(r) \approx B_{\mathrm{S}}\left\{
\begin{array}{l l l}
\left(\frac{R_\ast}{r}\right)^3, & \quad R_\ast\leq r<R_{\mathrm{A}}, \\
\frac{R_\ast^3}{R_{\mathrm{A}}r^2}, & \quad R_{\mathrm{A}} < r < R_{\mathrm{tor}}, \\
\frac{\upsilon_{\mathrm{rot}}}{\upsilon_\infty} \frac{R_\ast^2}{R_{\mathrm{A}}r}, & \quad R_{\mathrm{tor}}<r,
\end{array} \right.\]
where $R_{\mathrm{tor}}$ is the transition radius to a toroidal field defined by the ratation velocity of the star 
$\upsilon_{\mathrm{rot}}$ and the terminal velocity $\upsilon_\infty$ as $R_{\mathrm{tor}} = R_\ast (\upsilon_{\mathrm{rot}}/\upsilon_\infty)$. 
For LS 2883 $\upsilon_{\mathrm{rot}} \approx 450 $\,km\,s$^{-1}$ \citep{2011ApJ...732L..11N} and 
$\upsilon_\infty = 1350\pm200$\,km\,s$^{-1}$ \citep{1993AAS...183.1805M} implying 
$R_{\mathrm{tor}} \approx 3R_\ast$. In \psrb\ the pulsar is never coming this close 
to the Be star, therefore we can safely assume that the region where the cascades 
can efficiently develop is dominated by a magnetic field with a toroidal structure.
Moreover, when the pulsar is inside the disk at a large distance from the star ($\simeq45R_\ast\simeq2.9\times10^{13}$\,cm) 
and with a relatively short path within the disk \IStwo{($\simeq a_\mathrm{d, x} = 5\times10^{12}$)} the magnetic field 
can be assumed mono-directional.} \IStwo{Characteristic lengths of the \gaga-absorption and IC scattering are lower 
than the curvature radius of the magnetic field which is equal to the distance to the star. Mean free 
path of the \gaga-absorption process can be estimated as 
\begin{equation}
\lambda_{\gamma\gamma} = \left(\int n_{\mathrm{d}}(\nu) \sigma_{\gamma\gamma}(\epsilon_1,\epsilon_2) d\epsilon_2\right)^{-1},
\end{equation}
where $\epsilon_{\mathrm{i}} = h\nu_{\mathrm{i}}/m_\mathrm{e}c^2$ denotes normalized photon energy. Using the 
delta-function approximation of the \gaga\ cross section 
$\sigma_{\gamma\gamma} = (\sigma_\mathrm{T}/3) \epsilon_1 \delta(\epsilon_1 - 2/\epsilon_2)$ one can make 
a rough estimate of the mean free path for the primary photon with energy $E_{\gamma} = 100 E_{100}$\,GeV 
\begin{equation}
\lambda_{\gamma\gamma} (E_{\gamma}) \simeq 6.6\times10^{11} \left(\frac{A}{0.1}\right)^{-1} \left(E_{100}\right)^{3} \left(e^{3.3/E_{100}}-1\right)\,\mathrm{cm}.
\end{equation} 
Similarly, the characteristic length of the IC scattering in the Thomson regime can be estimated as 
\begin{equation}
\lambda_{\mathrm{IC}} \simeq 1.3\times10^{11} \left(\frac{A}{0.1}\right)^{-1}\,\mathrm{cm}.
\end{equation}
}

\IS{The strength of the surface magnetic field for Be stars is about $B_{\mathrm{S}} \sim 300 - 1000$\,G \citep{2012SSRv..166..145W}. 
For these values of $B_\mathrm{S}$ the Alfven radius $R_{\mathrm{A}}$ is equal to $(1-2)R_\ast$. Then 
at the distance of $45R_\ast$ the strength of the magnetic field is of the order of $1$\,G. 
To be on the safe side, we decided to consider a wider range of different values of magnetic field 
of $10^{-2} - 1$\,G in our calculations.  
}

\subsection{Numerical results}
\begin{figure}
\centering
\resizebox{\hsize}{!}{\includegraphics{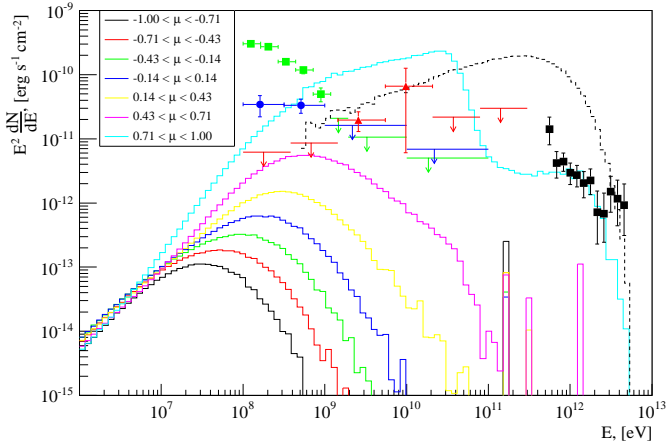}}
\caption{Cascade emission at different viewing angles $\theta = \arccos{\mu}$. Assumed parameters: $B_{x} = 0.01$\,G, $B_{y} = 0.001$\,G and 
$u_{\mathrm{d}} = 200$\,erg\,cm$^{-3}$. The black dashed line shows the input \gammaray\ spectrum. Black squares 
represent the 2011 H.E.S.S. data \citep{2013A&A...551A..94H}, blue circles \citep{2011ApJ...736L..11A} and red triangles \citep{2011ApJ...736L..10T} show the GeV 
data close to the 2010 periastron passage and green squares represent the GeV flare \citep{2011ApJ...736L..11A}.}
\label{cas_spec_ang}
\end{figure}
We used the cascade Monte Carlo code described above to calculate the angle-dependent cascade spectra for a variety 
of different input parameter sets within a parameter space motivated by the known properties of \IS{the Be star and its} 
circumstellar disk in the \psrb\ binary system. \IS{In these calculations we assume that the pulsar is a point-like source emitting 
isotropically in all directions. This assumption is reasonable if the wind termination shock at which leptons injected by the pulsar 
are efficiently accelerated and isotropized is close to the pulsar and the density of target seed photons is high enough to lead 
to IC scattering in the direct vicinity of the shock, which is the case for \psrb\ (see above). We approximate the spectrum of the 
VHE \gammarays\ as generated by IC scattering by electrons a power law distribution with an exponential cut-off. In our simulations we consider a 
mono-directional beam of photons to isolate all the geometrical effects. The source would emit the same beam of photons in every 
direction, and in the case of efficient cascading even those photons emitted in the opposite direction from the observer 
can contribute to the resulting observable spectrum. The orientation of the magnetic field would be different for primary photons 
emitted in different directions. Below we discuss the dependence of the cascade emission on the magnetic field orientation.
}

\begin{figure}
\centering
\resizebox{\hsize}{!}{\includegraphics{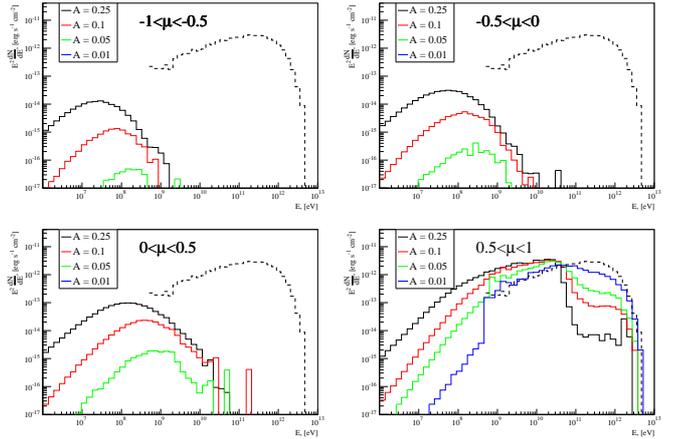}}
\caption{Dependence of the cascade emission on the energy density in the disk for different viewing angles. Coloured solid lines represent the spectra modified 
by gamma-gamma absorption and cascading for different assumed energy density as specified in the legend and the black dashed line shows the input \gammaray\ spectrum. 
The magnetic field is the same as in Fig. \ref{cas_spec_ang}. The location of the source is $x_{\mathrm{source}} = -0.5a_{\mathrm{d},\,x}$.}
\label{Adep}
\end{figure}

Figure \ref{cas_spec_ang} shows the viewing angle dependence of the cascade emission resulting from a monodirectional VHE $\gamma$-ray beam. 
For this simulation we chose a magnetic field of $B = 0.01$\,G oriented in a way that $B_{x} = 0.01$\,G 
and $B_{y} = 0.001$\,G, and an energy density of the target photon field of $u_{\mathrm{d}} = 200$\,erg\,cm$^{-3}$ ($A = 0.25$)\IS{, 
which corresponds to the upper limit for the disk energy density (see Section \ref{assump}).} 
The input \gammaray\ spectrum is described as an exponentially cut-off power law with a photon 
spectral index $\alpha = 1.5$ and a cut-off energy $E_{\mathrm{cut}} = 1$\, TeV. The resulting photon spectra for all 
directions were normalized with the same normalization factor, chosen to yield the observed TeV flux level of 
\psrb\ \citep{2013A&A...551A..94H}, as shown with black squares in Fig. \ref{cas_spec_ang}, in the forward direction. The input photon spectrum 
is shown as the black dashed line. For this high value of the energy density within 
the disk, gamma-ray-induced pair cascades would provide a significant contribution to the GeV $\gamma$-ray spectrum. 
However, the modified spectrum is in conflict with the GeV data obtained with \fermi\ \citep{2011ApJ...736L..11A, 2011ApJ...736L..10T}, 
which reveal very constraining upper limits in the same range of energies where the forward cascade emission peaks 
(cyan line in Fig. \ref{cas_spec_ang}). The comparison with the observed spectrum of the GeV emission detected at the periastron passage 
is not conclusive, because the spectral shape is quite controversial: two independent analyses revealed two different spectral shapes 
shown with blue circles \citep{2011ApJ...736L..11A} and red triangles \citep{2011ApJ...736L..10T} in Fig.\,\ref{cas_spec_ang}. This 
contradiction should be resolved by new observations around the periastron passage on 4th of May, 2014. Nevertheless, both spectral 
shapes provide upper limits in the energy range $20-100$ GeV, which are violated by the forward cascade emission. The spectrum of 
the GeV emission during the flaring period reveals a quite sharp cut-off consistent with an exponential cut-off at
$0.5$ GeV \citep{2013A&A...551A..94H} and upper limits in the energy range from 1 to 100 GeV. 
However, the GeV flare occurs when the pulsar is at the edge of the circumstellar disk or already 
after the pulsar has crossed the disk and, thus, no forward cascade emission is expected in the 
direction of the observer, while the \gammarays\ emitted in other directions 
can still reach the observer. The cascade emission in non-frontal directions peaks in the energy range 
in which the GeV flare emission is detected. Approximating \psrb\ as a point-source and taking into consideration that numerical calculations are 
provided only for a monodirectional beam of VHE $\gamma$-rays, the total observed cascade emission at GeV energies from the source can be estimated as 
the sum of all \gammarays\ emitted in non-frontal directions produced in cascades generated by a monodirectional beam, assuming that the 
cascade emission is not strongly dependent on the orientation of the primary beam. This cascade emission might be high enough to contribute 
significantly to the observed GeV flux. To study the possible contribution of the cascade emission 
to the observed GeV flux without violating \fermi\ upper limits at higher energies, we examine, in the following, the dependence of the cascade 
emission on the parameter $A$, i.e. energy density of the radiation field, on the strength and orientation of the magnetic field, and on the location 
of the pulsar within the radiation field.

\begin{figure}
\centering
\resizebox{\hsize}{!}{\includegraphics{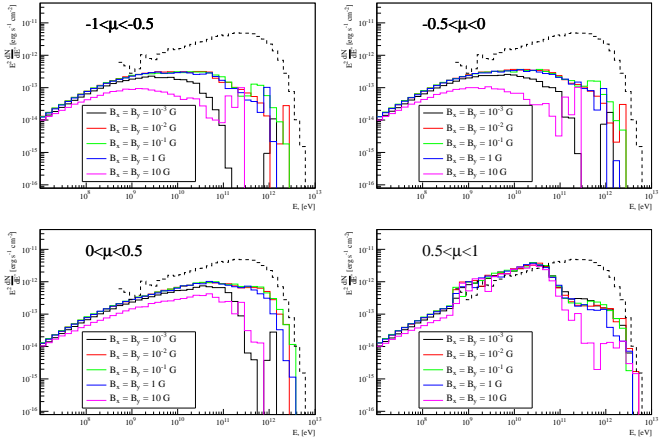}}
\caption{Dependence of the cascade emission on the magnetic field strength for different viewing angles. Coloured solid lines represent 
the spectra modified by gamma-gamma absorption and cascading for different field strength for a fixed agle of 45\deg\ between the direction 
of primary photons and magnetic field. Other assumed parameters: $A = 0.05$ and $x_{\mathrm{source}} = -0.5a_{\mathrm{d},\,x}$. 
The input \gammaray\ spectrum is shown with the black dashed line.}
\label{Bdep_strength}
\end{figure}

Figure \ref{Adep} illustrates the dependence on the energy density normalisation parameter $A$ for different values of 
$\mu = \cos\theta_{\rm obs}$, i.e. for the different viewing angles: the top left panel shows the spectra for $-1.0<\mu<-0.5$, the 
top right panel for $-0.5<\mu<0$, the bottom left panel for $0<\mu<0.5$, and the bottom right panel for $0.5<\mu<1.0$. 
For this simulation, we chose a
magnetic field of $B = 10^{-2}$\,G oriented in such a way that $B_{x} = 10^{-2}$\,G and $B_{y} = 10^{-3}$\,G,
and shifted the source location to the edge of the simulation box, along the $x$-axis at 
$x_{\mathrm{source}} = -0.5a_{\mathrm{d},\,x}$. The curves for the forward cascade emission (Fig. \ref{Adep}, bottom right panel) 
show that only for $A \lesssim 0.01$, i.e. $u_{\mathrm{d}}\lesssim 8$ erg/cm$^{3}$, the \fermi\ upper limits in the 
energy range $20-100$ GeV are not violated. For this low energy density, the cascade emission in other directions 
decreases dramatically (Fig. \ref{Adep}: left and top right panels) and will not contribute to the observable spectrum.

\begin{figure}
\centering
\resizebox{\hsize}{!}{\includegraphics{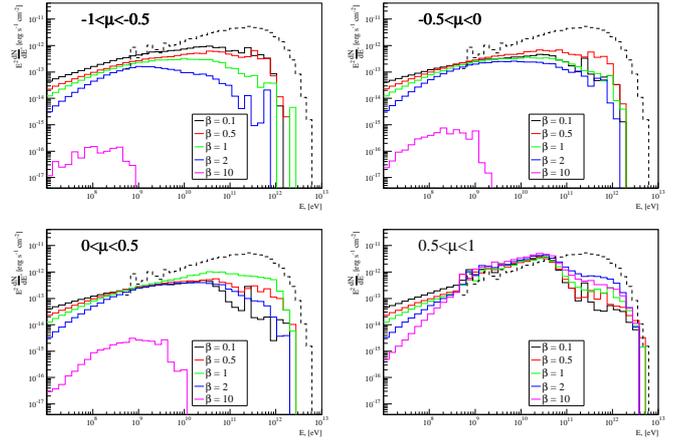}}
\caption{Dependence of the cascade emission on the magnetic field orientation for different viewing angles. Coloured solid lines represent 
the spectra modified by gamma-gamma absorption and cascading for different field orientation ($\beta = B_{\mathrm{x}}/B_{\mathrm{y}}$) for the same 
assumed magnetic field of $0.01$\,G. Other assumed parameters are the same as in Fig. \ref{Bdep_strength}. 
The input \gammaray\ spectrum is shown with the black dashed line.}
\label{Bdep_orient}
\end{figure}

Figures \ref{Bdep_strength} and \ref{Bdep_orient} show the dependence of the cascade emission on the strength and orientation 
of the magnetic field for different viewing angles (arranged in the same way as in Fig.\,\ref{Adep}). 
For this simulation we assumed $A = 0.05$ and $x_{\mathrm{source}} = -0.5a_{\mathrm{d},\,x}$. 
The cascade emission is strongly dependent on both the strength and the orientation of the magnetic field. 
For the same orientation, the stronger magnetic field increases the cascade emission in the non-frontal 
directions (Fig. \ref{Bdep_strength} top-left, top-right and bottom-left panels). For stronger magnetic fields, 
electrons and positrons are effectively isotropised before they produce secondary \gammarays\ through 
the IC scattering. For the magnetic field strength of $\gtrsim 0.01$\,G electrons and positrons are 
completely isotropised and thus already the first generation of cascade photons is completely isotropised 
and emitted evenly in all directions. \IS{For the magnetic field strength of $\gtrsim10$\,G, the magnetic field 
energy density becomes comparable with the photon energy density (in this case $u_{\mathrm{B}}/u_{\mathrm{d}}\gtrsim0.2$) 
which results in significant synchrotron losses.} 

To study the dependence of the cascade emission on the orientation of the magnetic field, we introduce the parameter 
$\beta = B_{\mathrm{x}}/B_{\mathrm{y}}$ as the ratio of the $x$- and $y$-components of the magnetic field. 
For the same strength of the magnetic field of $0.01$\,G, a higher value of $\beta$, i.e. a stronger $x$-component results 
in a shift of the peak of the non-frontal cascade emission to lower energies. However, this is coincident with a 
decrease of the flux level of the non-frontal cascade emission. A "co-alignment" of the magnetic field with 
the direction of the primary photon beam ($\beta \gg 1$) reduces the contribution of the non-frontal 
cascade emission dramatically, since almost all the secondary photons are emitted in the forward direction. 
For $\beta \leq 1$, i.e. for a magnetic field oriented at large angle with respect to the primary photon direction, the cascade emission 
in all directions peaks at roughly 30 GeV. The lower the value of $\beta$, the more cascade radiation is emitted 
backwards (Fig. \ref{Bdep_orient} top-left panel).

\begin{figure}
\centering
\resizebox{\hsize}{!}{\includegraphics{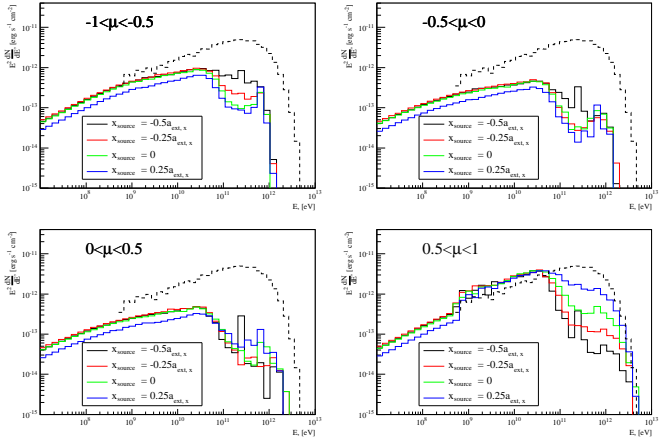}}
\caption{Dependence of the cascade emission on the source location along the $x$-axis for different viewing angles. Coloured solid lines represent 
the spectra modified by gamma-gamma absorption and cascading for different values of $x_{\mathrm{source}}$. Assumed parameters: $B_{x} = 10^{-3}$\,G, $B_{y} = 10^{-2}$\,G and $A = 0.05$. 
The input \gammaray\ spectrum is shown with the black dashed line.}
\label{locdep}
\end{figure}

The dependence of the cascade emission on the location of the source within the disk for 
different viewing angles (same as in Fig. \ref{Adep}) is shown in Fig. \ref{locdep}. 
For this simulation, we chose a magnetic field of $B = 10^{-2}$\,G oriented in such a way that $B_{x} = 10^{-3}$\,G 
and $B_{y} = 10^{-2}$\,G, and $A = 0.05$. The bottom-right panel of Fig.\,\ref{locdep} confirms that --- as expected --- the level of the 
\gaga-absorption of the \gammaray\ beam directed along the $x$-axis decreases as the source is moving towards 
the front edge of the radiation field region. However, for $x_{\mathrm{source}} \leq  0$ the cascade emission at the energies below $\sim 50$\,GeV 
does not depend significantly on the location of the source. 
This reflects the fact that in such configurations, a high-opacity regime ($\tau_{\gamma\gamma} \gtrsim 1$) has been reached, and a further increase of the opacity only has a minor effect on the $\gamma\gamma$ absorption and cascading process. 
Minor differences still occur only in the energy range $50$\,GeV$-1$\,TeV and 
only in the backward directions (Fig.\,\ref{locdep} top panels). This can be naturally explained by the immediate escape of 
first generation cascade photons emitted in backward directions as the source is located close to the back edge of the 
radiation field region. Obviously, the effect of the location of the source inside the radiation field region 
depends on the value of the energy density as the $\gamma\gamma$ opacity is proportional to the energy density for a given radiation temperature.

\newref{\section{Discussion}}
\newref{\subsection{GeV emission}}
The \fermi\ data taken around the 2010 periastron passage were analysed by two independent groups 
\citep{2011ApJ...736L..11A, 2011ApJ...736L..10T} yielding similar results for the flaring 
period, but quite different results for the first detection period close to periastron, when 
a very low GeV flux was observed. Both light curves and spectral shapes of the emission for this period 
obtained by the two different groups are considerably different. \citet{2011ApJ...736L..11A} announced the first 
detection of the source integrating the data in the so-called "brightening" period from $t_{\mathrm{p}}-28$~d to 
$t_{\mathrm{p}}+18$~d. The light curve study on a weekly basis showed a significant flux in the periods $t_{\mathrm{p}}-19$~d to 
$t_{\mathrm{p}}-12$~d and $t_{\mathrm{p}}+2$~d to $t_{\mathrm{p}}+16$~d with an upper limit between those periods. However, a subsequent 
re-analysis of the same data by the same group showed a constant flux in the period from $t_{\mathrm{p}}-25$~d to 
$t_{\mathrm{p}}+16$~d \citep{2014MNRAS.439..432C}. In contrast, \citet{2011ApJ...736L..10T} detected the low GeV flux 
in the period from about a month before periastron to periastron with no significant emission after periastron until 
the GeV flare.

There is also an essential discrepancy in the obtained spectra for the "low-flux" periods for two different groups 
as shown in Fig. \ref{Adep}. These two spectra contradict each other revealing a significant flux in different energy 
bands ($0.1 - 1$\,GeV \citep{2011ApJ...736L..11A} and $1 - 20$\,GeV \citep{2011ApJ...736L..10T}). Resolving this 
spectral descrapancy is very important in terms of comparison of the data to the simulated cascade emission and 
drawing any conclusions. The spectrum obtained by \citet{2011ApJ...736L..11A} (blue circles in Fig. \ref{Adep}) 
rejects the possibility of a significant cascade contribution, because even the slightest absorption of the TeV 
\gammarays\ and re-emission at lower energies would violate the \fermi\ upper limit in $1-100$\,GeV energy 
band. Therefore, the \fermi\ upper limit in this case constrains the energy density in the disk to 
$u_{\mathrm{d}}\lesssim 8$ erg/cm$^{3}$, i.e. $A\lesssim0.01$.

\begin{figure}
\centering
\resizebox{\hsize}{!}{\includegraphics{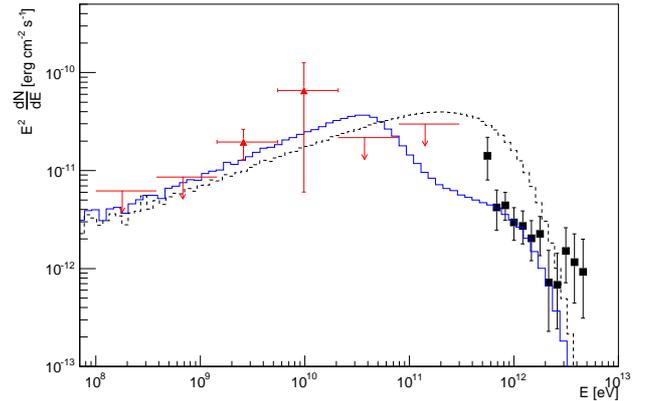}}
\caption{Comparison with the data. The blue solid line represents the spectrum of the forward emission ($0.98\leq\mu\leq1$) 
modified by gamma-gamma absorption and cascading. Assumed parameters: $B = 0.1$\,G, $\beta = 10$, $u_{\mathrm{d}} = 20$ erg/cm$^{3}$, 
and $x_{\mathrm{source}} = -0.5a_{\mathrm{d},x}$. Black squares represent the 2011 H.E.S.S. data \citep{2013A&A...551A..94H} and red triangles 
show the GeV data close to the 2010 periastron passage as reported by \citet{2011ApJ...736L..10T}.}
\label{cas_data}
\end{figure}

In contrast, the spectrum presented by \citep{2011ApJ...736L..10T} is consistent with a substantial contribution from cascade 
emission. The energy of the peak in the spectrum of the cascade \gammarays\ emitted in the forward direction 
only slightly depends on the energy density and magnetic field (see Figures \ref{Adep}, \ref{Bdep_strength} and 
\ref{Bdep_orient}) and is in the energy range of $15-30$ GeV. Figure\,\ref{cas_data} shows the comparison of the 
GeV-TeV data with the simulation results of the cascade Compton radiation \IS{from the primary photon beam directed towards the observer} 
emitted in the forward direction into a cone with an opening angle 
11\deg\ ($0.98\leq\mu\leq1$). This corresponds to the cascade emission radiated in the direction of the observer from the primary 
\gammarays\ emitted within the cone of the same opening angle assuming that the magnetic field affects the cascades initiated by every primary photon 
identically. This assumption is valid for sufficiently small values of the opening angle, such that the orientation 
of the magnetic field with respect to the direction of the primary photon does not change much and its influence on 
the shape of the cascade emission spectrum is negligible. The simulation was performed for $A = 0.025$, i.e. 
$u_{\mathrm{d}} = 20$ erg/cm$^{3}$, magnetic field $B = 0.1$ G with $\beta = 10$ and the location of the 
source $x_{\mathrm{source}} = -0.5a_{\mathrm{d},x}$. We decided to choose the magnetic field which is oriented close to 
the direction of the primary photon \IS{(and in the direction of the observer)} since in this case  
we avoid the dependence on the orientation, because secondary 
photons are emitted in the same direction as the primary photon. Otherwise, the cascade radiation emitted in other directions 
can become higher than the forward emission and a cone with a bigger opening angle should be considered, which in turn requires an accurate 
calculation of the magnetic field orientation for every primary photon. \IS{This assumption, however, is close to the realistic scenario. 
The direction of the toroidal magnetic field in the disk region is close to the direction towards the observer for given low values 
of the orbital inclination angle of about $23$\deg\ \citep{2011ApJ...732L..11N} and the disk inclination angle with respect to the orbit 
of about $10$\deg.}

The disappearance of the GeV emission in the period from 
16 to 30 days after periastron can be naturally explained within this model by the fact that the forward cascade emission 
decreases as the pulsar moves towards the frontal edge of the disk. In the framework of this model, the GeV 
emission should appear during the first crossing of the disk and stay constant until the second crossing of the disk. 
This is consistent with the light curve reported by \citet{2011ApJ...736L..11A}, but does not agree with the light 
curve reported by \citet{2011ApJ...736L..10T}, which reveals the disappearance of the GeV emission after periastron. 
Therefore, the proposed model of the non-flaring GeV emission from the system is consistent with the spectrum 
reported by \citet{2011ApJ...736L..10T} and light curve reported by \citet{2011ApJ...736L..11A}, but inconsistent 
vice versa.

\IS{It should be noted though that simulations presented above are performed considering only disk radiation field and 
neglecting the stellar radiation field. As discussed in Section \ref{assump} this assumption can be valid when the pulsar is located 
in the disk environment where the energy density of the disk photons is higher than the energy density of stellar photons. But if the 
pulsar is close to periastron the energy density of the stellar radiation is higher and the efficient cascading is possible before 
\gammarays\ reach the disk environment. The modeling of the cascade emission from the pulsar at periastron would require 
to take into account the stellar photons as well as the toroidal structure of the magnetic field, since the path of the 
\gammaray\ within the system is long and the direction of the magnetic field will change considerably. Therefore, the comparison of the model with 
the "periastron" emission presented on Fig. \ref{cas_data} should be treated only as a hint that the GeV emission may be 
explained by the cascade radiation (even if we consider only disk photons). This implies that taking into account stellar 
photons might probably provide a better fit of the data and decrease the estimate (upper limit) of the disk energy density. A detailed study 
of the cascades development in the system which would properly account for all the effects discussed in this work is in progress 
and will be applied to the new GeV data taken around 2014 periastron.} \newref{Preliminary analysis reported in \citet{2014ATel.6198....1T} 
revealed no significant GeV emission, i.e. $>5\sigma$, close to the 2014 periastron, with a 90\% confidence level upper limit about $2-3$ times 
lower than the flux reported around the previous periastron passage. This result, if confirmed, might provide evidence for periastron-to-periastron 
variability.}

The GeV flare is most probably generated by some other process for several reasons. Although the cascade emission radiated in the 
backward direction may contribute to the flux at $0.1-1$\,GeV, this contribution is small compared to the GeV flare even for 
very high values of the energy density. Moreover, for high values of the energy density in the disk we expect a strong cascade emission 
in the forward direction at energies $1-100$\,GeV. Although during the GeV flare, when the pulsar is escaping the disk, no forward 
cascade emission is expected, it should have been observed in earlier phases. Finally, if the flare is due to cascade emission, 
the same effect should be observed before and at the beginning of the first crossing of the disk, but this is not seen.

\newref{\subsection{TeV light curve}}

\begin{figure}
\centering
\resizebox{\hsize}{!}{\includegraphics{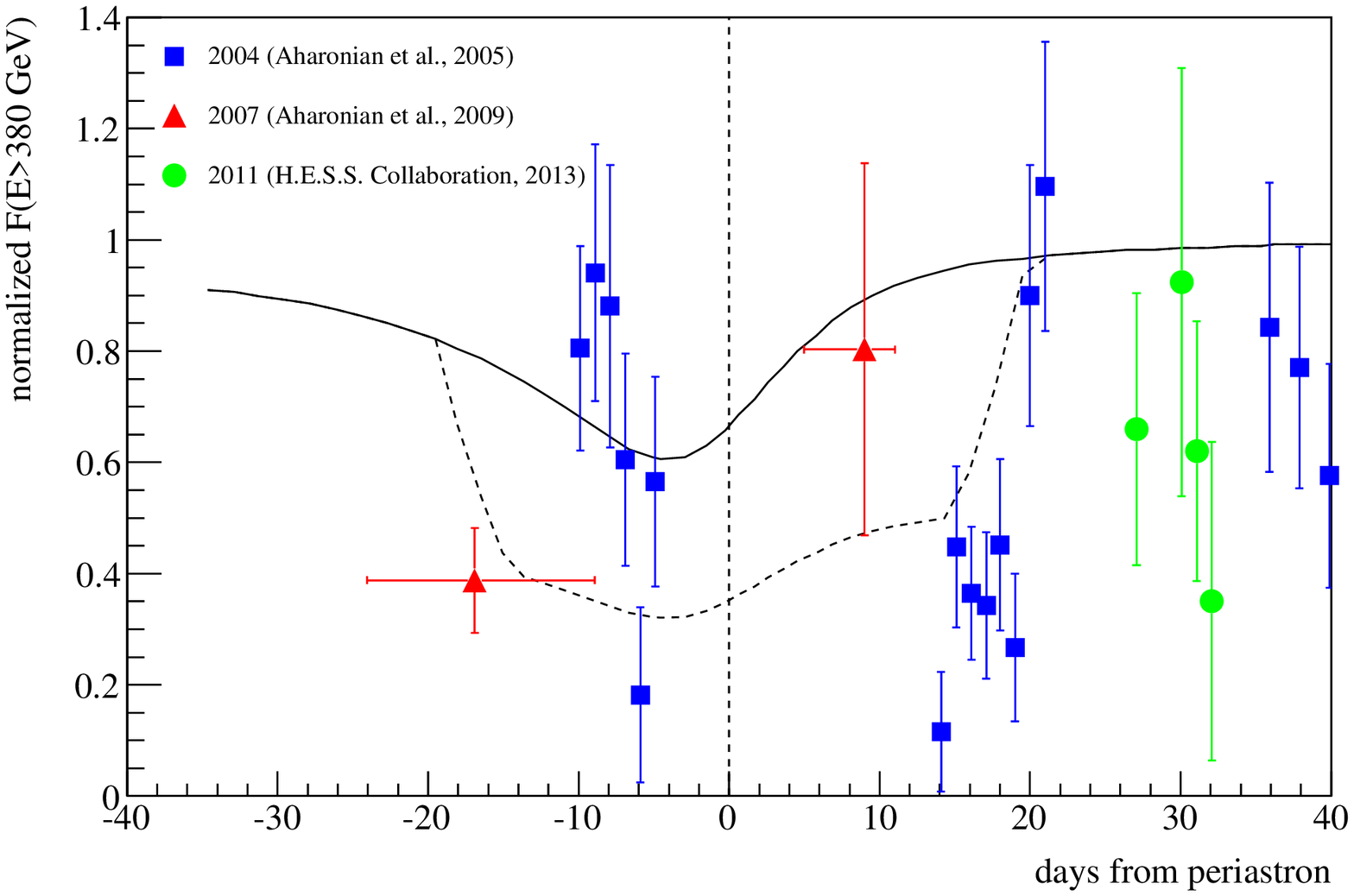}}
\caption{\newref{VHE $\gamma$-ray flux modulation caused by \gaga-absorption in \psrb. The solid line represents the normalized integrated flux above 380 GeV 
after absorption caused by the stellar radiation field as calculated in \citet{2006A&A...451....9D}. The dashed line shows the same flux 
taking into account also the absorption caused by the disk radiation field. 
See the main text for details on the geometry and other parameters.
The data points represent the combined H.E.S.S. 
light curve from three observed periastron passages as reported in \citet{2013A&A...551A..94H}. The integrated flux above 1\,TeV is extrapolated 
down to 380\,GeV using a photon index of $2.8$. The same flux normalisation factor as in \citet{2006A&A...451....9D} of $10^{-11}$\,cm$^{-2}$\,s$^{-1}$ 
is used.}}
\label{tev_lc}
\end{figure}

\newref{The absorption of the \gammaray\ emission caused by the interaction with the disk radiation field might have 
a significant impact at the observed TeV light curve. \citet{2006A&A...451....9D} calculated the resulting integrated 
flux above 380\,GeV after absorption caused by the stellar radiation field. It was shown that the absorption due to the 
stellar radiation plays only a minor role in the observed variability. However, the disk radiation was not considered in those 
calculations. Using the estimate of the resulting integrated flux after the absorption on stellar photons only, $F_{\mathrm{s}}$, 
calculated in \citet{2006A&A...451....9D} we calculated the resulting flux after the total absorption due to both stellar and disk 
radiation fields as
\begin{equation}
F(E>380\,\mathrm{GeV}) = \frac{F_{\mathrm{s}}(E>380\,\mathrm{GeV})}{\epsilon_{\gamma,\,\mathrm{max}} - \epsilon_{\gamma,\,\mathrm{min}}} \int_{\epsilon_{\gamma,\,\mathrm{min}}}^{\epsilon_{\gamma,\,\mathrm{max}}}e^{-\tau_{\gamma\gamma,\,\mathrm{d}}(\epsilon_\gamma)} d\epsilon_{\gamma},
\end{equation}
where $\tau_{\gamma\gamma,\,\mathrm{d}}(\epsilon)$ is the optical depth of the disk radiation field, $\epsilon_\gamma = E_\gamma/(m_{\mathrm{e}}c^2)$ is the normalised \gammaray\ energy and the integration is performed between 
$E_{\gamma,\,\mathrm{min}} = 380$\,GeV and $E_{\gamma,\,\mathrm{max}} = 10$\,TeV. For these calculations we used the same orbital parameters as used 
in \citet{2006A&A...451....9D}: orbital period $P_{\mathrm{orb}} = 1236.7$, eccentricity $e = 0.87$, periastron longitude $\omega = 138.7^\circ$, 
and inclination angle $i = 35^\circ$. The disk inclination angle is assumed to be $i_{\mathrm{d}}=10^{\circ}$. The disk is assumed to be 
perpendicular to the major axis of the orbit with a constant width of $10^{12}$\,cm (corresponds to $1^\circ$ half-opening angle at a 
distance of $45R_\ast$) and constant energy density of $8$\,erg\,cm$^{-3}$ (the highest energy density for which the \fermi\ upper limits are 
not violated, see previous subsection). The expected light curve after absorption assuming a constant initial \gammaray\ flux is shown 
in Figure \ref{tev_lc}. The solid line represents the flux after absorption taking into account only stellar photons as reported 
by \citet{2006A&A...451....9D} and the dashed line shows the flux after the total absorption including the disk radiation field. It can be seen 
that the circumstellar disk might play a crucial role in the flux variability around the periastron passage and that the observed flux variability 
can be explained by the absorption. The time shift of the observed preperiastron peak with respect to the modelled light curve might be explained 
by the rotation of the disk normal with respect to the major axis. The assumption of the constant width of the disk results in an overestimation 
of the absorption close to periastron, since the disk should be thiner close to the star. However, the energy density in the disk is expected to be highest close to the star. The combined effect might result in a shallower minimum near periastron than predicted by our calculations. 
The light curve has been modelled under the assumption that the primary \gammaray\ 
flux is constant. This is a valid approximation only in the case when IC scattering occurs in the saturation regime which is believed to be true around 
periastron. Farther from periastron the primary flux would decrease. In the case when IC scattering is not in the saturation regime, local 
maxima are expected at the disk crossings and at the periastron. This would completely change the expected light curve. With the current quality 
of the observed TeV light curve it is difficult to draw any final conclusions. New observations conducted in 2014 might bring some 
clarity and give answers to some of these open questions. }

A critical test of the \newref{absorption} model could be derived from the study of the spectral variability of the TeV emission. The TeV spectrum 
is expected to soften when the pulsar is moving through the disk for the first time, then the spectral shape should remain 
\newref{almost} unchanged \newref{with a slight variability due to the absorption in the stellar radiation field} until 
the pulsar moves out of the disk after periastron. After that the spectrum should harden again. Unfortunately, the VHE $\gamma$-ray 
observations during previous periastron passages did not provide enough statistics to draw firm conclusions about the spectral variability. 
Although there are some indications of 
spectral changes both during the 2004 \citep{2005A&A...442....1A} and the 2007 \citep{2009A&A...507..389A} periastron passages,  these 
changes are not significant. However, if the spectral variability is real on both occasions, it suggests a hardening of the spectrum 
towards periastron, which is in contradiction with the effect we expect in the case of effective cascading.

\section{Summary}
\label{summary}
We investigated the generation of \gammaray\ induced pair cascades 
in the circumstellar disk of the companion Be star in the binary system \psrb\ and the possible 
contribution of the cascade emission to the observed \gammaray\ spectrum of the source. It is shown that 
the cascade emission generated in the disk cannot be responsible for the GeV flare 
observed about 30 days after periastron. However, the cascade emission might explain the GeV emission 
observed close to periastron. The spectrum and the light curve of the GeV emission close to periastron 
are not well resolved (two different groups obtained results which contradict each other), which is probably 
because of a very low flux in this period of time. This uncertainity in the light curve and spectral shape widen 
a scope of different mechanisms which can explain the GeV emission. We showed that the cascade emission generated 
in the disk may make a non-negligible contribution to this GeV emission. \newref{We also showed that the absorption 
of the \gammaray\ emission in the disk might explain the observed TeV light curve.} Observations around the 2014 
periastron passage might provide further tests to \newref{our model} \IS{and serve as a guideline to a more 
detailed modeling.} 



{
\vspace{0.5in}
\normalsize
\noindent
{\bf References}
}

\bibliographystyle{elsarticle-harv} 
\bibliography{references.bib}


%% \begin{thebibliography}{00}


%% \bibitem[ ()]{}

%% \end{thebibliography}
\end{document}